\documentclass[twocolumn,showpacs,preprintnumbers,amsmath,amssymb]{revtex4}

\usepackage{graphicx}
\usepackage{dcolumn}
\usepackage{bm}

\begin{document}

\title{EL2--like defects in InP nanowires}

\author{ R.   H.  Miwa, and T. M. Schmidt}

\affiliation{Instituto de  F\'{\i}sica, Universidade  Federal de
Uberl\^andia,  C.P.  593,  38400-902, Uberl\^andia,  MG -  Brazil} 

\author{A. Fazzio}

\affiliation{Instituto de F\'{\i}sica, Universidade de S\~ao Paulo,
 Brazil}
 
\date\today

\begin{abstract}

 We  have performed  an {\it  ab initio}  total  energy investigation,
 within the  density functional theory  (DFT), of antisite  defects in
 InP nanowires  (InP$NW$s) grown along the [111]  direction. Our total
 energy results indicate that, (i)  P antisites (P$_{\rm In}$) are the
 most  likely antisite  defect  compared with  In antisites  (In$_{\rm
 P}$), and  (ii) the formation energies  of P and In  antisites do not
 depend on the $NW$ diameter.  In particular, thin InP$NW$s, diameters
 of  $\sim$13~\AA,  the  P$_{\rm  In}$ antisite  exhibits  a  trigonal
 symmetry,  lying at  0.15~\AA\ from  the  $T_d$ site,  followed by  a
 metastable  configuration  with   P$_{\rm  In}$  in  an  interstitial
 position (1.15~\AA\ from  the $T_d$ site). We find  a P$_{\rm In}$--P
 dissociation energy of 0.33~eV, and  there is no EL2--like center for
 such a thin InP$NW$.   However, EL2--like defects occur by increasing
 the  $NW$  diameter.   For  diameters of  $\sim$18~\AA,  the  P$_{\rm
 In}$--P dissociation  energy increases  to 0.53~eV, which  is 0.34~eV
 lower compared  with the P$_{\rm In}$--P dissociation  energy for InP
 bulk  phase,  0.87~eV.   We   mapped  the  atomic  displacements  and
 calculated the  relaxation energy, Franck--Condon  shift, upon single
 excitation of P$_{\rm In}$  induced states in InP$NW$.  The formation
 (or not)  of EL2--like  defects, P$_{\rm In}$--P  dissociation energy
 barrier, and  the Franck--Condon (FC)  energy shift, can be  tuned by
 the $NW$ diameter.

\end{abstract} 


\maketitle

\section{Introduction}

Antisite is one of the most studied native defect in III--V compounds, for
instance, the EL2 center in GaAs.  In a seminal work, Kami\'nska et
al.~\cite{KaminskaPRL1985} established that the EL2 center exhibits a
tetrahedral symmetry, formed by an isolated As antisite defect (As$_{\rm
  Ga}$).  Afterward, several
experimental~\cite{bardelebenPRB1986,kabirajAPL2005} as well as theoretical
studies~\cite{chadiPRL1988,DabrowskPRL1988,ZhangPRB1993,chadiPRB2003,OvehofPRB2005}
have been done aiming to clarify the structural and electronic properties of
EL2 defect in GaAs.  After more than ten year of investigations, currently
there is a general agreement that the stable EL2 center is ruled by an
isolated As$_{\rm Ga}$ with $T_d$ symmetry, while the metastable EL2$^M$
structure is attributed to the As$_{\rm Ga}$ atom displaced along the $C_{3v}$
axis.

Its  is  well  known  that  low temperature  growth  of  other  III--V
semiconductors,  not  only  GaAs,  allows the  preparation  of  highly
nonstoichiometric compounds.   Indeed, anion antisite  defects in InP,
P$_{\rm  In}$, has been  identified for  the first  time unambiguously
through the  electron paramagnetic resonance technique  by Kennedy and
Wilsey~\cite{kenedyAPL1984}.   Dreszer  et  al.~\cite{dreszerPRB1993},
using  Hall,  high--pressure  far--infrared absorption  and  optically
detected  magnetic   resonance  measurements  (at   low  temperature),
verified that  InP has two  dominant donor levels associated  with the
phosphorus antisite defect. Further investigations, based upon {\it ab
initio} total  energy calculations, proposed the  formation of P$_{\rm
In}$  antisites clusters in  InP bulk~\cite{tmsrhmPRB1999}.   

The most  interesting property  on this class  of defects is  the fact
that      they      also      exhibit     a      clear      metastable
behavior~\cite{chadiPRL1988,mikuckiPRB2000,CaldasPRL1990}.   A neutral
anion--antisite defect in III--V compounds has a stable fourfold and a
metastable    threefold   interstitial   configuration,    where   the
anion--antisite      is     displaced      along      the     $C_{3v}$
axis~\cite{CaldasPRL1990},  similarly  to   EL2  and  EL2$^M$  defects
observed in GaAs.  This center can be photo--excited into a metastable
state,  and from  which  it returns  to  the ground  state by  thermal
activation.  Thus,  it is expected a  persistent photoconductivity for
P$_{\rm In}$ at low temperature.

Nowadays the electronic properties of several materials can be tailored
throughout the manipulation/control of their size in the atomic scale. Within
this new class of (nano)structured materials, the semiconductors nanowires
($NW$s) are attracting great deal of interest for future applications in
several types of nanodevices.  In particular, InP-nanowires (InP$NW$s) have
been considered as a potential structure for fabrication of sensors, light
emitting diodes, and field effect
transistors~\cite{jianfangSci2001,xiangfengNat2001}.  Usually the
vapor--liquid--solid mechanism, with a gold particle seed, has been utilized
for the growth of these nanostructures~\cite{bhuniaAPL2003}.  These materials
are quasi--one--dimensional with electrons confined perpendicularly to the
$NW$ growth direction.  They exhibit interesting electronic and optical
properties due to quantum confinement effects, viz.: the size dependence of
InP$NW$ bandgap~\cite{hengNatMat2003,TSchmidtPRB2005}.  Recent {\it ab initio}
calculations, performed by Li et al.~\cite{LiPRL2005}, suggested the formation
of stable $DX$ center in small GaAs quantum--dots,  dot diameter smaller
than $\sim$15~nm.

It is quite likely the formation of native defects during the $NW$ growth
process.  Therefore, the structural and electronic properties of those defects
are important issues to be addressed, in order to improve our understanding of
native defects in quasi--1D semiconducting $NW$ systems.

In  this  paper  we  carried  out  an {\it  ab  initio}  total  energy
investigation  of antisite  defects  in InP$NW$s.   We  find that  the
formation of EL2--like defects,  antisite dissociation energy, and the
Franck--Condon (FC)  energy are  ruled by the  $NW$ diameter.   On the
other  hand, the antisite  formation energies  are insensitive  to the
$NW$  diameter, being P$_{\rm  In}$ antisites  the most  likely defect
compared with In$_{\rm P}$.   At the equilibrium geometry, the P$_{\rm
In}$  atom   in  thin  InP$NW$  (diamenter  of   13~\AA)  exhibits  an
energetically  stable  trigonal  symmetry,  followed by  a  metastable
configuration with P$_{\rm In}$ in an interstitial position, 1.15~\AA\
from the $T_d$  site.  In this case, the  P$_{\rm In}$--P dissociation
energy     is    0.33~eV.      Increasing     the    $NW$     diameter
(13~$\rightarrow$~18~\AA),  we  find  a P$_{\rm  In}$--P  dissociation
energy  of  0.53~eV,  where  P$_{\rm  In}$ lying  on  the  $T_d$  site
represents the  energetically most stable configuration.   For the InP
bulk  phase,  the P$_{\rm  In}$--P  dissociation  energy  is equal  to
0.87~eV.  Similarly to the bulk phase, P$_{\rm In}$ in InP$NW$ induces
the formation of localized states within the energy bandgap.  However,
for such a thin $NW$ system, this defect does not exhibit an EL2--like
behavior. On  the other hand, increasing the  $NW$ diameter, EL2--like
defects are expected to  occur.  Finally, within a constrained density
functional   approach~\cite{ArtachoPRL2004},   we   map   the   atomic
displacements  along thin  InP$NW$ upon  single excitation  of P$_{\rm
In}$ induced  states, and calculate the  respective relaxation energy,
``Franck-Condon'' (FC) shift.

\section{Theoretical Approach}

Our  calculations  were performed  in  the  framework  of the  density
functional theory  (DFT)~\cite{kohn}, within the  generalized gradient
approximation  due to  Perdew, Burke,  and  Ernzerhof~\cite{PBE}.  The
electron--ion interaction was  treated by using norm--conserving, {\it
ab   initio},  fully   separable   pseudopotentials  \cite{KL}.    The
Kohn--Sham  wave   functions  were   expanded  in  a   combination  of
pseudoatomic numerical orbitals~\cite{sankey}.   Double zeta basis set
including polarization  functions (DZP)  was employed to  describe the
valence  electrons~\cite{dzp}.    The  self--consistent  total  charge
density  was obtained  by  using the  SIESTA code~\cite{siesta}.   The
InP$NW$  was modeled  within  the supercell  approach,  where the  InP
bilayers  were piled  up along  the [111]  direction  with periodicity
length of $2\sqrt 3 a$ and diameters of 13 and 18~\AA\ ($a$ represents
the optimized lattice constant  along the [111] direction of InP$NW$).
The $NW$ surface dangling bonds were saturated with hydrogen atoms.  A
mesh cutoff of 170~Ry was  used for the reciprocal--space expansion of
the total charge density, and  the Brillouin zone was sampled by using
one special  {\bf k}  point. We have  verified the convergence  of our
results with respect  to the number and choice of  the special {\bf k}
points.  All atoms  of the nanowire were fully  relaxed within a force
convergence criterion of 20 meV/\AA.

\begin{figure}[h]
\includegraphics[width= 7cm]{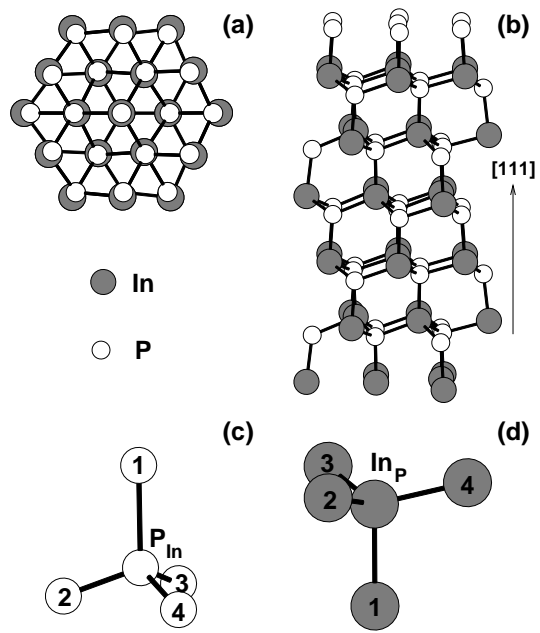}
\caption{Structural models  of thin InP$NW$, diamenter  of 13~\AA. (a)
Top view and  (b) side view.  The atomic  geometry around the antisite
defects are indicated in (c) P$_{\rm In}$, and (d) In$_{\rm P}$.}
\label{side}
\end{figure}

\section{Results and Comments}

\begin{figure}[h]
\includegraphics[width= 8cm]{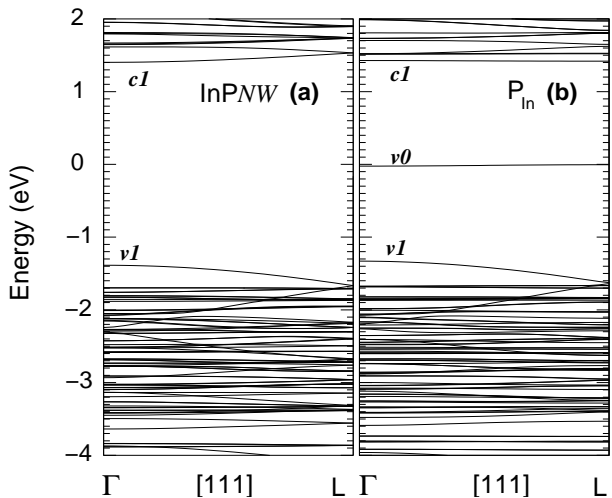}
\caption{Electronic band  structures of (a) perfect  thin InP$NW$, and
(b)  defective thin  InP$NW$ with  a P  antisite (P$_{\rm  In}$). $NW$
diameter of 13~\AA.}
\label{Bandas2}
\end{figure}

Figure~\ref{side}  presents  the  atomic  structure of  thin  InP$NW$,
diameter  of  13~\AA,  growth  along  the [111]  direction,  top  view
[Fig.~\ref{side}(a)] and side  view [Fig.~\ref{side}(b)] (the hydrogen
atoms   are  not  shown).    Due  to   the  1D   quantum  confinement,
perpendicularly  to  the $NW$  growth  direction,  the  energy gap  of
InP$NW$    increases   compared   with    the   bulk    InP   (1.0~eV)
\cite{TSchmidtPRB2005}.   We find energy  gaps of  2.8 and  2.2~eV for
$NW$ diameters  of 13  and 18~\AA, respectively.   It is  important to
take into  account those energy gaps are  underestimated, with respect
to   the   experimental  measurements,   within   the  DFT   approach.
Figure~\ref{Bandas2}(a)  presents electronic  band  structure of  thin
InP$NW$ for  wave vectors parallel  to the [111]  direction ($\Gamma$L
direction). At the $\Gamma$ point, the highest occupied states exhibit
an energy split of 0.31 and  0.20~eV ($a_1 + e$), for $NW$ diameter of
13 and 18~\AA,  respectively. The valence band maximum  of bulk InP is
described   by   a   three--fold   degenerated  $t_2$   state.    Such
($t_2\rightarrow   a_1  +  e$)   energy  splitting   is  due   to  the
$T_d\rightarrow C_{3v}$  symmetry lowering of InP$NW$  with respect to
the bulk phase.

The energetic  stability of antisites  in InP$NW$s can be  examined by
the  calculation of  formation energies  ($\Omega_i$).   The formation
energy of  P and In  antisites, P$_{\rm In}$  [Fig.~\ref{side}(c)] and
In$_{\rm P}$ [Fig.~\ref{side}(d)], respectively, can be written as,
$$ \Omega_i =  E[{\rm InP}NW_i] - E[{\rm InP}NW]  - n_{\rm In}\mu_{\rm
In} - n_{\rm P}\mu_{\rm P}.
$$

We have considered the formation of antisites inner InP$NW$s.  $E[{\rm
InP}NW_i]$ represents the  total energy of InP$NW$ with  $i$ = P$_{\rm
In}$  or In$_{\rm  P}$ antisite  defect, and  $E[{\rm InP}NW]$  is the
total energy of a perfect InP$NW$.  $n_{\rm In}$ ($n_{\rm P}$) denotes
the number of  In (P) atoms in  excess or in deficiency. The  In and P
chemical potentials,  $\mu_{\rm In}$ and  $\mu_{\rm P}$, respectively,
are  constrained  by  following thermodynamic  equilibrium  condition,
$\mu_{\rm In}  + \mu_{\rm P} = \mu_{\rm  InP}^{Bulk}$, where $\mu_{\rm
InP}^{Bulk}$ is the chemical potential  of bulk InP.  Under In rich (P
poor)  condition  we   will  have  $\mu_{\rm  In}\rightarrow  \mu_{\rm
In}^{Bulk}$,  whereas  under In  poor  (P  rich) condition,  $\mu_{\rm
In}\rightarrow     \mu_{\rm    In}^{Bulk}     -     \Delta    H_f({\rm
InP})$~\cite{qianPRB1988}.   For the  heat of  formation of  bulk InP,
$\Delta  H_f(\rm InP)$,  we  have considered  its experimental  value,
0.92~eV.

\begin{figure}[h]
\includegraphics[width= 7cm]{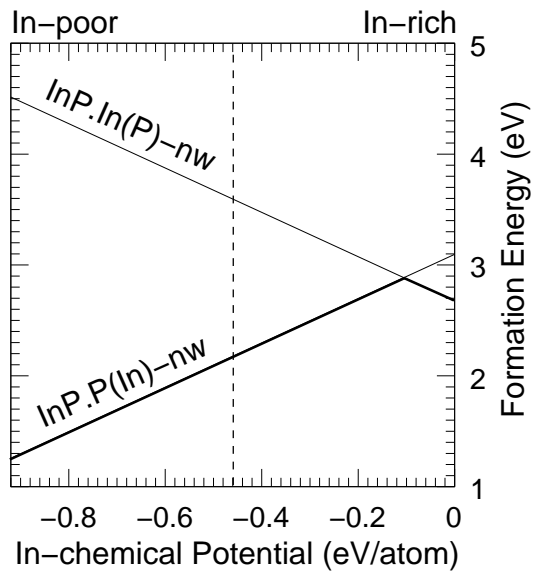}
\caption{Formation  energies   of  P$_{\rm  In}$   and  In$_{\rm  P}$
antisites in  thin InP$NW$,  diameter of 13~\AA,  as a function  of In
chemical potential.}
\label{FormEnergy}
\end{figure}

Figure~\ref{FormEnergy} presents our  calculated results of $\Omega_i$
for  thin  InP$NW$ (diameter  of  13~\AA), as  a  function  of the  In
chemical potential.  It is clear that the formation of P$_{\rm In}$ is
dominant compared  with In$_{\rm P}$.  This latter  defect occurs only
for  In rich  condition.  At  the  In and  P stoichiometric  condition
(dashed  line  in   Fig.~\ref{FormEnergy})  we  obtained  $\Omega_{\rm
P_{In}}$ = 2.15~eV and  $\Omega_{\rm In_P}$ = 3.57~eV.  Increasing the
$NW$  diameter  (18~\AA), we  find  $\Omega_{\rm  P_{In}}$ =  2.18~eV.
Within the  same calculation procedure, we  obtained similar formation
energy results for  the InP bulk phase, viz.:  $\Omega_{\rm P_{In}}$ =
2.14~eV and  $\Omega_{\rm In_P}$  = 3.59~eV.  Those  results indicates
that, (i) there is an  energetic preference of P$_{\rm In}$ antisites,
compared with In$_{\rm  P}$, for both structural phases  of InP.  (ii)
The  formation energy  of $\rm  P_{In}$ does  not depend  on  the $NW$
diameter.

\begin{figure}[h]
\includegraphics[width= 8cm]{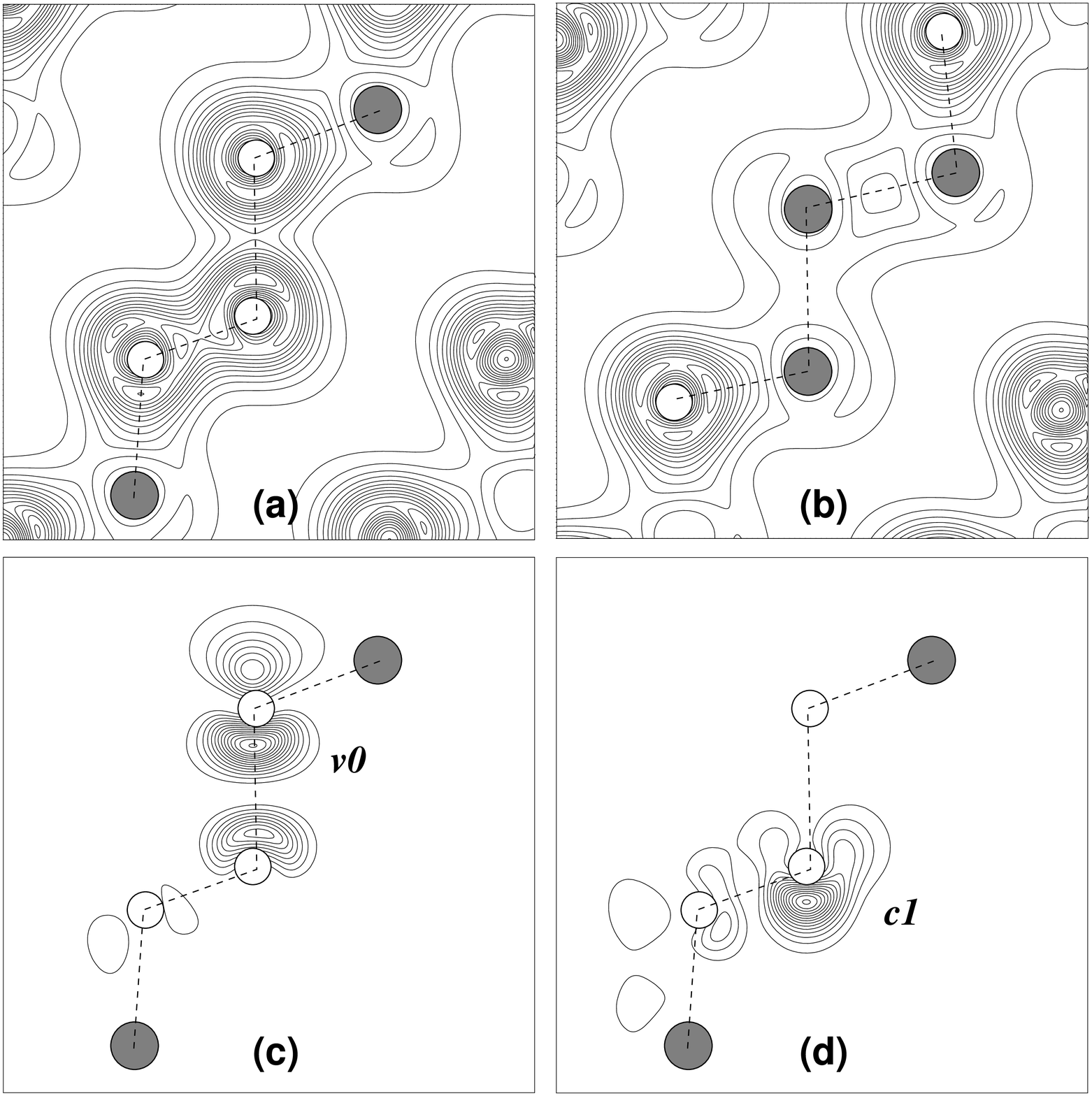}
\caption{Total charge densities of (a) P$_{\rm In}$, and (b) In$_{\rm P}$
  antisites in InP$NW$. The partial charge densities of P$_{\rm In}$ induced
  (c) highest occupied $v0$ and (d) lowest unoccupied $c1$ states.}
\label{rhoTotParc2}
\end{figure}

At the equilibrium geometry, P$_{\rm In}$ defect in bulk InP keeps the
$T_d$ symmetry  with P$_{\rm In}$--P  bond lengths of  2.49~\AA, while
In$_{\rm P}$  exhibits a weak Jahn--Teller distortion  along the [001]
direction.   Those results  are in  accordance with  previous  {\it ab
initio}        studies        of        antisites        in        InP
\cite{seitsonenPRB1994,tmsrhmPRB1999,castletonPRB2004}.   On the other
hand, the equilibrium  geometry of antisites in thin  InP$NW$ is quite
different.  The P$_{\rm In}$--P$_1$  bond, parallel to the $NW$ growth
direction, is stretched by 27\%  compared with the other three P$_{\rm
In}$--P$_i$ bonds,  $i=2-4$ in Fig.~\ref{side}(c).   Similarly for the
In$_{\rm  P}$ antisites,  In$_{\rm P}$--In$_1$  is stretched  by 8.6\%
compared  with the  other three  In$_{\rm P}$--In  bonds  indicated in
Fig.~\ref{side}(d).           Figures~\ref{rhoTotParc2}(a)         and
\ref{rhoTotParc2}(b)  depict  the  total  charge densities  along  the
P$_{\rm In}$--P$_1$  and In$_{\rm P}$--In$_1$  bonds, respectively. In
particular,  due to  the  large P$_{\rm  In}$--P$_1$ bond  stretching,
P$_{\rm  In}$ is  weakly bonded  to P$_1$,  whereas the  other P$_{\rm
In}$--P$_2$, --P$_3$ and --P$_4$ bonds [see Fig.~\ref{rhoTotParc2}(a)]
exhibit a  strong covalent character.  Therefore,  different from bulk
InP,  P$_{\rm  In}$  antisites  in  $NW$ system  exhibits  a  $C_{3v}$
symmetry,  with the  P$_{\rm In}$  atom displaced  from $T_d$  site by
0.15~\AA\  along the [111]  axis.  Such  a P$_{\rm  In}$ displacement,
along the $C_{3v}$ axis, has  not been observed by increasing the $NW$
diameter to  18~\AA.  In this case,  the P$_{\rm In}$  atom accupies a
$T_d$ site and the P$_{\rm In}$--P bond lengths are equal to 2.94~\AA.

\begin{figure}[h]
  \includegraphics[width= 7cm]{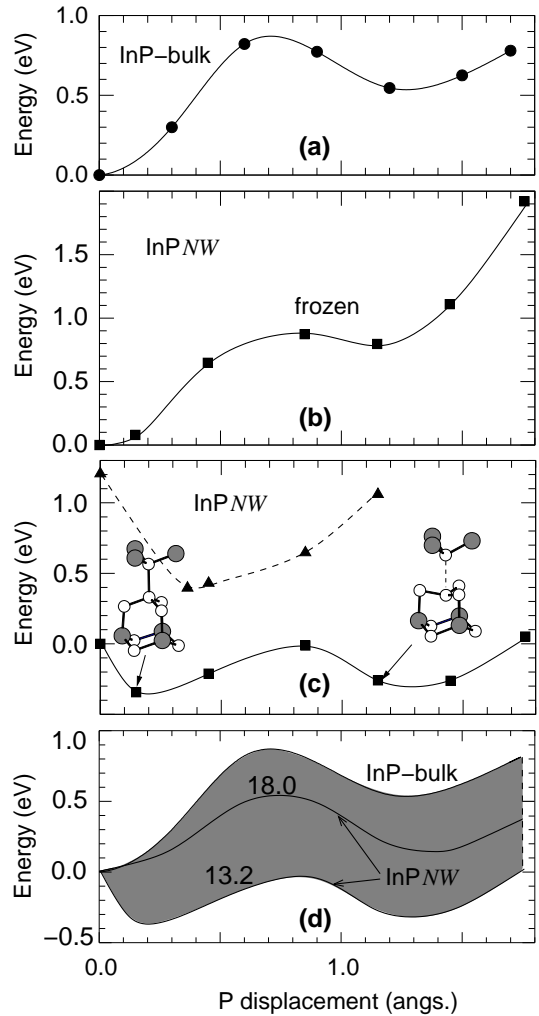}
\caption{Total Energy as a function of the P$_{\rm In}$ antisite 
  displacement along the [111] direction: (a) bulk InP, and (b) InP$NW$ where
  the In and P atoms are not allowed to relax during the P$_{\rm In}$
  displacement. (c) Whole In and P atomic position along the $NW$ are allowed
  to relax. In (c) the total energy barrier indicated by triangles (dashed
  line) was calculated for an excited electronic configuration, single
  excitation. (d) Energy barriers for bulk InP and thin InP$NW$ systems.}
\label{DisplEnergy2}
\end{figure}

Several  anion antisite defects  in III--V  materials were  studied by
Caldas et al.~\cite{CaldasPRL1990}.  In  that work, based upon {\it ab
initio}   total   energy  calculations,   the   authors  observed   an
energetically stable $T_d$ symmetry  for P$_{\rm In}$ and a metastable
$C_{3v}$   configuration   for   the   antisite  atom   displaced   by
$\sim$1.3~\AA\    along    the    [111]   direction.     Indeed,    in
Fig.~\ref{DisplEnergy2}(a) we  present our  total energy results  as a
function of P$_{\rm In}$  displacement, along the [111] direction, for
the InP  bulk phase.  We find  that the $T_d$  symmetry represents the
energetically most stable configuration, followed by an energy barrier
of 0.87~eV at 0.7~\AA\ from the  $T_d$ site, z = 0.7~\AA, breaking the
P$_{\rm   In}$--P$_1$   bond.    Finally,  the   metastable   $C_{3v}$
configuration occurs for z =  1.2~\AA, where the P$_{\rm In}$ occupies
an     interstitial    site.      Figures~\ref{DisplEnergy2}(b)    and
\ref{DisplEnergy2}(c) present  the energy barrier for  P$_{\rm In}$ in
thin InP$NW$.   We have examined  two different processes: (i)  the In
and P  atoms of the $NW$ are  not allowed to relax  during the P$_{\rm
In}$  displacement  [Fig.~\ref{DisplEnergy2}(b)]. (ii)  The  In and  P
coordinates are  fully relaxed  for each P$_{\rm  In}$ step  along the
[111]       direction,      i.e.       an       adiabatic      process
[Fig.~\ref{DisplEnergy2}(c)].   It  is   noticeable  that  the  energy
barrier calculated in (i) is very similar to that obtained for P$_{\rm
In}$   in   bulk  InP   [Fig.~\ref{DisplEnergy2}(a)].    In  (i)   the
energetically most  stable configuration  for P$_{\rm In}$  exhibits a
$T_d$ symmetry, and  we find a dissociation energy of  0.84~eV for z =
0.86~\AA.   Further  P$_{\rm In}$  displacement  indicates a  $C_{3v}$
metastable geometry  for z $\approx$ 1.15~\AA, where  the P$_{\rm In}$
atom occupies  an interstitial site.  Meanwhile, in  (ii) the $C_{3v}$
symmetry, with the P$_{\rm In}$ atom lying at 0.15~\AA\ from the $T_d$
site,  represents the  energetically most  stable  configuration.  The
P$_{\rm  In}$--P$_1$ dissociation energy  reduces to  0.33~eV (at  z =
0.85~\AA),  and  there  is  a  metastable  geometry  for  z  $\approx$
1.15~\AA.  The  total energy difference  between the stable  ($S$) and
the  metastable ($M$)  configurations, $E(S)  - E(M)$,  in  InP$NW$ is
equal   to  $-$0.084~eV,   while  in   bulk  InP   we  find   $E(S)  -
E(M)$~=~$-$0.53~eV.  Comparing (i) and  (ii) we verify that the atomic
relaxation plays a fundamental rule not only to the energy barrier for
the P$_{\rm  In}$ atomic displacement  along the [111]  direction, but
also  for the  equilibrium  geometry of  the  P$_{\rm In}$  structure.
Increasing  the $NW$  diameter, the  energy barrier  for  P$_{\rm In}$
displacement will  become similar  to that obtained  for bulk  InP (an
upper     limit    for    very     large    $NW$     diameter),    see
Fig.~\ref{DisplEnergy2}(d).  Indeed,  for $NW$ diameter  of 18~\AA, we
find a P$_{\rm In}$--P$_1$ dissociation energy of 0.53~eV, and $E(S) -
E(M)$ equal  to $-$0.14~eV.  Those  results indicate that  the P$_{\rm
In}$--P$_1$ dissociation  barrier, and the $E(S) -  E(M)$ total energy
difference,  can  be  tuned  within  the shaded  region  indicated  in
Fig.~\ref{DisplEnergy2}(d).

Focusing on  the electronic structure,  the formation of  P$_{\rm In}$
defect  in thin  InP$NW$  gives rise  to  a deep  double donor  state,
labeled  $v0$ in  Fig.~\ref{Bandas2}(b), lying  at 1.3~eV  above $v1$.
Such  a  P$_{\rm In}$  induced  state  is  very localized  within  the
fundamental band gap, being almost flat along the $\Gamma$L direction.
For P$_{\rm  In}$ in  bulk InP, {\it  ab initio} studies  performed by
Seitsonen et al.~\cite{seitsonenPRB1994} indicates the formation of an
occupied state at 0.7~eV above  the valence band maximum.  The partial
charge density contour  plot of $v0$, Fig.~\ref{rhoTotParc2}(c), shows
an  anti-bonding  $\pi^\ast$ orbital  concentrated  along the  P$_{\rm
In}$--P$_1$  bond.   The  lowest  unoccupied  state,   $c1$,  is  also
concentrated   on  the   P$_{\rm   In}$  antisite,   as  depicted   in
Fig.~\ref{rhoTotParc2}(d).

EL2 and EL2$^M$ centers in III--V are characterized by a stable $T_d$ and a
metastable $C_{3v}$ geometries for an isolated V$_{\rm III}$ interstitial
atom.  However, before the V$_{\rm III}$ antisite arrives to the metastable
$C_{3v}$ configuration, there is a strong electronic coupling between the
highest occupied and the lowest unoccupied antisite induced states near to the
local maximum for the V$_{\rm III}$--V dissociation energy, see Fig.~3 in
Ref.~\cite{CaldasPRL1990} and Fig.~4 in Ref.~\cite{DabrowskPRB1989}.

\begin{figure}[h]
\includegraphics[width= 7cm]{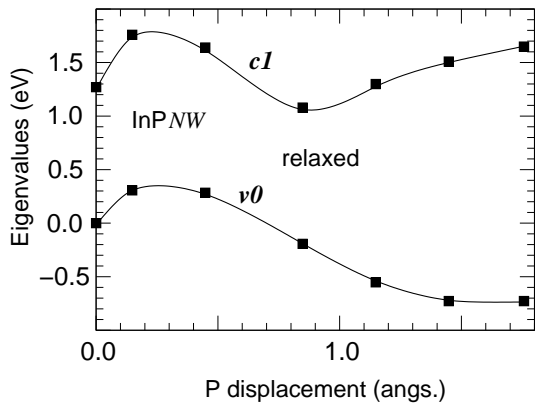}
\caption{Single particle eigenvalues ($\epsilon$) for P$_{\rm In}$: 
  highest occupied $v0$ [Fig.~\ref{rhoTotParc2}(c)] and lowest unoccupied $c1$
  [Fig.~\ref{rhoTotParc2}(d)].}
\label{eigenvalues}
\end{figure}

Similarly,  for thin  InP$NW$,  we examined  the  energy positions  of
single particle  eigenvalues, $\epsilon(v0)$ and  $\epsilon(c1)$, as a
function of the P$_{\rm In}$ displacement along the $C_{3v}$ axis. Our
calculated  results, depicted  in Fig.~\ref{eigenvalues},  reveal that
there is  no electronic  coupling between $v0$  and $c1$.  Thus, based
upon the  calculated energy barrier  [Fig.~\ref{DisplEnergy2}(c)], and
the evolution  of the antisite induced  states, $v0$ and  $c1$, we can
state that there is no EL2--like center in thin InP$NW$s. On the other
hand,  increasing   the  $NW$  diameter  [Fig.~\ref{DisplEnergy2}(d)],
EL2--like center is expected to occur in InP.

Within the  DFT approach, we have examined  the structural relaxations
upon single excitation  from the highest occupied state  ($v0$) to the
lowest unoccupied  state ($c1$) of  P$_{\rm In}$ in thin  InP$NW$.  We
have used the calculation procedure  proposed by Artacho et al., where
the single excitation was modeled  by ``promoting an electron from the
highest  occupied molecular  orbital (HOMO)  to the  lowest unoccupied
molecular     orbital    (LUMO)''\cite{ArtachoPRL2004},     i.e.     a
``constrained'' DFT calculation. So  that, promoting one electron from
$v0$ to $c1$, $v0\rightarrow c1$, we obtained the equilibrium geometry
(full relaxations)  as well as the  self--consistent electronic charge
density.    The   atomic   relaxations   along  thin   InP$NW$,   upon
$v0\rightarrow  c1$  single  excitation,  are localized  near  to  the
P$_{\rm In}$  position.  Since we  are removing one electron  from the
anti-bonding $\pi^\ast$ orbital depicted in Fig.~\ref{rhoTotParc2}(c),
the P$_{\rm In}$--P$_1$ bond  shrinks by 0.12~\AA\ (2.90 $\rightarrow$
2.78~\AA), and there is radial  a contraction of $\sim$0.2~\AA\ at z =
6.8~\AA.

The  structural relaxations  are  proportional to  the degradation  of
optical energy due to the  atomic displacements associated with such a
$v0 \rightarrow c1$ single excitation, FC shift.  Comparing the radial
and the longitudinal displacements, we  verify that the most of the FC
shift comes  from the  atomic relaxation parallel  to the  $NW$ growth
direction.   The  atomic  displacements  along  the  radial  direction
($\rm\Delta  r$) are localized  nearby P$_{\rm  In}$, lying  within an
interval  of $-0.2<\Delta  r<0$~\AA.  While  the  atomic displacements
parallel to the growth direction ($\rm\Delta z$) lie within a range of
$-0.2<\Delta  z<0.3~$\AA, being less  concentrated around  the P$_{\rm
In}$ antisite position.  We find a FC energy shift of 0.8~eV, which is
quite  large  compared with  the  InP  bulk  phase.  Within  the  same
calculation approach,  we find  a FC energy  shift of 0.05~eV  for InP
bulk.  Previous {\it  ab initio} study indicates a  FC energy shift of
0.1~eV for bulk  InP.  Thus, suggesting that the  FC energy shift also
can be tuned by controlling the $NW$ diameter.

\section{Conclusions}

In  summary,  we  have  performed  an {\it  ab  initio}  total  energy
investigation of antisites  in InP$NW$. We find that  the P$_{\rm In}$
antisite is the  most likely to occur.  For  thin InP$NW$s the P$_{\rm
In}$ atom  exhibits a  trigonal symmetry, followed  by a  metastable P
interstitial  configuration.  Increasing  the $NW$  diameter (18~\AA),
P$_{\rm In}$  occupying the $T_d$ site becomes  the energetically most
stable configuration.  The calculated energy barrier, for P$_{\rm In}$
displacement  along the  $C_{3v}$  axis, indicates  that  there is  no
EL2--like  defect  in thin  InP$NW$s.   However,  increasing the  $NW$
diameter,  13  $\rightarrow$  18~\AA,  we  observe  the  formation  of
EL2--like defects  in InP.  Within a ``constrained''  DFT approach, we
calculated  the structural relaxations  and the  FC energy  shift upon
single excitation  of the electronic  states induced by  P$_{\rm In}$.
We inferred that not only the formation (or not) of EL2--like defects,
but also (i)  the P$_{\rm In}$--P dissociation energy,  (ii) the total
energy   difference  between  stable   and  metastable   P$_{\rm  In}$
configurations,  $E(S)-E(M)$, and  the (iii)  FC energy  shift  can be
tuned by controlling the InP$NW$ diameter.

\begin{center}
  {\large\bf Acknowledgments}
\end{center}

The authors acknowledge financial support from the Brazilian agencies CNPq,
FAPEMIG, and FAPESP. The most of calculations were performed using the
computational facilities of the Centro Nacional de Processamento de Alto
Desempenho/CENAPAD-Campinas.

\bibliography{../RHMiwa}

\end{document}